\documentclass[twocolumn,showpacs,preprintnumbers,amsmath,amssymb]{revtex4}
\newcommand{\myskip}[1]{}
\newcommand{\K}{{\cal K}}

\newcommand{\gex}{{\rm ge}}

\newcommand{\ct}{c_1}
\newcommand{\cts}{c_1^2}

\newcommand{\q}{{\bf q}}

\newcommand{\half}{\frac{1}{2}}

\renewcommand{\d}{{\rm d}}

\newcommand{\p}{\partial}

\newcommand{\BEQ}{\begin{eqnarray}}
\newcommand{\EEQ}{\end{eqnarray}}
\newcommand{\BEA}{\begin{eqnarray}}
\newcommand{\EEA}{\end{eqnarray}}
\newcommand{\nn}{\nonumber }
\renewcommand{\d}{{\rm d}}

\newcommand{\mn}{{\mu\nu}}

\renewcommand{\u}{\rho c^2}

\def\dbarrm {{\mathchar'26\mkern-11mu{\rm d}}}                       %
\usepackage{graphicx}% Include figure files
\usepackage{dcolumn}% Align table columns on decimal point
\usepackage{bm}% bold math

\begin{document}

%\preprint{ITFA 777}

\title{Note on the chemical potential of %quasi relativistic 
decoupled matter in the Universe} %neutrinos}

\author{ Th. M. Nieuwenhuizen$^{1,2}$ and C. Pombo$^3$}

%\email{t.m.nieuwenhuizen@uva.nl}

% \homepage{http://staff.science.uva.nl/~nieuwenh/}
\affiliation{
$^{1}$Institute for Theoretical Physics, University of Amsterdam, Science Park 904, P.O. Box 94485, Amsterdam, the Netherlands\\
$^{2}$ Center for Cosmology and Particle Physics, New York University,  4 Washington Place, New York, NY 10003, USA \\
$^3$ 1 Washington Square Village, New York, NY  10012, USA}
%Eikenweg 2, 1092 BN Amsterdam, The Netherlands}

%\date{\today}

\begin{abstract} 
Textbooks on cosmology exhibit a thermodynamic inconsistency for free streaming, decoupled matter.
It is connected here to the chemical potential, which deviates from its equilibrium value
$\mu=\alpha k_BT$, where $\alpha$ is the usual parameter of the Fermi-Dirac or Bose-Einstein distribution function.
\end{abstract}

%\pacs{04.70.Bw}{Classical black holes}
%\pacs{04.20.Cv}{Fundamental problems and general formalism}
%\pacs{04.20.Jb}{Exact solutions}

\maketitle

%\begin{multicols}

%\documentclass[doublecol]{epl2} \renewcommand{\cal }{\rm } \renewcommand{\mb }{ {\overline{ m}} } \newcommand{\half}{\frac{1}{2}} \renewcommand{\subsubsection}{\subsection} $\surd16\qquad$

\newcommand{\beh}{{\rm BEH}}

\section{Motivation}

In the textbooks on cosmology and astrophysics, such as the ones  of Dodelson \cite{Dodelson}, Weinberg \cite{Weinberg} 
and of Giunti and Kim \cite{GiuntiKim}, chapters on thermodynamics 
state  that the pressure $p$ and  energy density $u=\rho c^2$ ($\rho$ is the mass density, $c$ the vacuum speed of light in vacuum) 
satisfy at temperature $T$ the relation

\BEQ \label{thermoprho}
\frac{\d p}{\d T}=\frac{u+p}{T}.
\EEQ
Dodelson mentions that in principle there occurs also the chemical potential, adding that it is usually irrelevant in cosmology \cite{Doddel}.
These authors all rush to point out that (\ref{thermoprho}) is satisfied for photons and other relativistic particles, 
where $u\sim T^4$ and $p=\frac{1}{3}u$, but  they do not mention that it is  violated in other cases.
Indeed, for nonrelativistic matter, such as baryons, cold dark matter and non-relativisitic neutrinos, 
where $u=nm c^2$, $p\ll u$, with $n$ the particle density, this relation is violated.
% Because there is no doubt on the correctness of the ensuing physics, 
We seek here to explain the cause of this paradox.

Our main point will be that because thermodynamics should be valid beyond equilibrium, an old thruth
that has allowed to formulate thermodynamics for the glassy state~\cite{LeuzziN}, 
Eq. (\ref{dpdTdef}) below must substitute Eq. (\ref{thermoprho}).
Hereto we recall the relevant laws of thermodynamics, apply them to equilibrium and, next, to the
out of equilibrium situation in the expanding universe.

\section{Thermodynamics}

The first law states that the change of energy  of the system, $U=\u V$, is the sum of the heat added to
the system, the work done on the system and energy gained by adding particles to the system,

\BEQ \d U=\dbarrm Q+\dbarrm W+\mu\d N. 
\EEQ
where $\mu$ is the gain in energy per particle when one keeps $\dbarrm Q=\dbarrm W=0$.
With the second law $\dbarrm Q=T\d S$ and the mechanical work for a change of the volume $V$,
$\dbarrm W=-p\d V$ this yields 

\BEQ \label{dS=}
\d S=\frac{\d U+p\d V-\mu\d N}{T}.
\EEQ
The system is extensive, implying that $S(\lambda U,\lambda V,\lambda N)=\lambda S(U,V,N)$.
Therefore it must hold  that

%\BEQ \label{S123=}S=s_1\left(\frac{U}{V},\frac{N}{V}\right)U
%+s_2\left(\frac{U}{V},\frac{N}{V}\right)V
%+s_3\left(\frac{U}{V},\frac{N}{V}\right)N,\EEQ

\BEQ \label{S123=}S=s_1(\rho,n)U+s_2(\rho,n)V+s_3(\rho,n)N,
\EEQ
for certain functions $s_{1,2,3}$.
% of the mass density $\rho=U/Vc^2$ and the number density $n=N/V$.
% and $T$,  where the energy density $\rho$ and particle density $n$ are defined in the usual way,
%\BEQ\rho=\frac{U}{V},\qquad n=\frac{N}{V}.\EEQ
Considering $\d S$ from (\ref{S123=}) and comparing its $\d U$, $\d V$ and $\d N$ terms with (\ref{dS=}),
it is seen that $s_1=1/T$, $s_2=p/T$, $s_3=-\mu/T$, which %ere $T$, $p$ and $\mu$
are functions of $\rho$ and $n$. These results imply the Euler relation

\BEQ \label{S=}
S=\frac{U+pV-\mu N}{T},
\EEQ
or, taken per unit volume,
\BEQ \label{s=}
s=\frac{u+p-\mu n}{T}.
\EEQ

Consistency between (\ref{dS=}) and (\ref{S=}) imposes also

%\BEQ \label{dp=}V\d p=%S \d T+N\d\mu %=\frac{U+pV-N\mu}{T}\d T+N\d\mu.\EEQ
\BEQ \label{dp=}
\d p=\frac{\rho+p-\mu n}{T}\d T+n \d\mu.
\EEQ
Inverting the relation $T=T(\rho,n)$ at  fixed $n$, one obtains $\rho=\rho(T,n)$, which expresses
all  these quantities as functions of $T$ and $n$. This allows to derive from (\ref{dp=}) the relations
\BEQ
\label{pppT=}
\left.\frac{\p p}{\p T}\right| _n=\frac{u+p-\mu n}{T}+\left.n\frac{ \p\mu}{\p T}\right|_n
\EEQ
and

\BEQ
\label{pppn=}
\left.\frac{\p p}{\p n}\right| _T=\left.n\frac{ \p\mu}{\p n}\right|_T.
\EEQ
While $n$ is fixed for cooling a gas in a fixed, closed volume, we can also admit situations where $n$ 
changes with $T$, as it happens when cooling a gas with a movable piston or in the expanding universe.
Then we are interested in the total derivative

\BEQ \label{dpdTdef}
\frac{\d p}{\d T}=\left.\frac{\p p}{\p T}\right| _n
+\left.\frac{\p p}{\p n}\right| _T\frac{ \d n}{\d T},
\EEQ
and a similar definition for $\d \mu/\d T$. Combining (\ref{pppT=}), (\ref{pppn=}), (\ref{s=}) and (\ref{dpdTdef}) we obtain
\BEQ \label{dpdT=}
\frac{\d p}{\d T}=\frac{u+p-n\mu}{T}+n\frac{\d\mu}{\d T}=s+n\frac{\d\mu}{\d T},
\EEQ
a result that could also have been obtained directly by dividing 
(\ref{dp=}) by $\d T$. This expresses Eq. (\ref{pppT=}) for  the general case.
From (\ref{S=}) we have the thermodynamic relation

\BEQ\label{mu=}
 \mu=\frac{u+p-Ts}{n}.
 \EEQ

\section{Equilibrium thermodynamics of ideal quantum gases}

In thermal equilibrium the grand canonical distribution function
of an ideal Fermi-Dirac gas with mass $m$ and chemical potential $\mu=\alpha k_BT$
reads for a mode labeled by ${\bf q}$ and having energy $E(\q)$,

\BEQ \sum_{n_q=0}^1e^{n_\q[\alpha-\beta E(\q)]}=1+e^{\alpha-\beta E(\q)}\EEQ
where $\beta=1/k_B T$. Accounting for all modes yields

\BEQ {\cal Z}=\prod_\q\left(1+e^{\alpha-\beta E(\q)}\right)
\EEQ
%the Bose-Einstein and 
The Fermi-Dirac distribution reads

\BEQ f(\q)=\frac{1}{e^{\beta E(\q)-\alpha}+1}.
\EEQ

We are interested in an ideal gas of particle with mass $m$ and momentum $q$.
Taking periodic boundary conditions for a cube with size $V^{1/3}$, 
the allowed momenta are $(q_x,q_y,q_z)=2\pi\hbar V^{-1/3}(n_x,n_y,n_z)$, with integer values of
$n_{x,y,z}$, so that

\BEQ E(q)=\sqrt{m^2c^4+q^2c^2},
\EEQ
and

\BEQ \log{\cal Z}=V\int\frac{\d^3q}{(2\pi\hbar)^3}\log\left(1+e^{\alpha-\beta E(q)}\right).
\EEQ
From this we derive

\BEQ \label{dlogZ=} \d\log{\cal Z}=
\frac{\d V}{V}\log{\cal Z}+N\d\alpha-U\d\beta,
\EEQ
in which we may identify $U=Vu$ and $N=Vn$ from

\BEQ\label{n=} n=\int\frac{\d^3q}{(2\pi\hbar)^3}f,
\EEQ
\BEQ\label{u=}
\qquad u=\int\frac{\d^3q}{(2\pi\hbar)^3}fE,\qquad
\EEQ
and conclude that
\BEQ %\log{\cal Z}=\frac{pV}{k_BT},\qquad 
p=nk_BT \log{\cal Z}.
\EEQ
Reinserting this into (\ref{dlogZ=}) we verify the central relation (\ref{dpdT=}).
We finally find from (\ref{s=}) that the entropy density equals

\BEQ \label{s(f)=}
s=k_B\int\frac{\d^3q}{(2\pi\hbar)^3}[-f\log f-(1-f)\log(1-f)],
\EEQ
To derive this result we may use that
\BEQ \beta u-\alpha n=
\int\frac{\d^3q\,f(\beta E-\alpha)}{(2\pi\hbar)^3}=
\int\frac{\d^3q}{(2\pi\hbar)^3}f\log\frac{1-f}{f}, \nn
\EEQ
and 

\BEA
\beta p&=&\int\frac{\d^3q}{(2\pi\hbar)^3}\log\left(1+e^{\alpha-\beta E}\right)\nn\\& =&
\int\frac{\d^3q}{(2\pi\hbar)^3}\log\frac{1}{1-f}.
\label{pold}
\EEA
For later use we point out that by inserting $\sum_{i=1}^3\p{ q_i}/{\p{ q_i}}=3$ in the first
identity and performing a partial integration, we obtain the equivalent results
\BEA
p&=&-\frac{T}{3}\int\frac{\d^3q}{(2\pi\hbar)^3}{\bf q}\cdot\frac{\p}{\p{\bf q}}\log\left(1+e^{\alpha-\beta E}\right)\nn\\
& =&
\int\frac{\d^3q}{(2\pi\hbar)^3}\, f\,\frac{q^2c^2}{3E}.
\label{pnew}
\EEA

We can also verify from (\ref{mu=}) that $\alpha$ takes the equilibrium value
\BEQ 
\alpha=\beta \mu.
\EEQ

For bosons one has the grand canonical partition sum 

\BEQ {\cal Z}=\prod_\q\left(1-e^{\alpha-\beta E(\q)}\right)^{-1},
\EEQ
the Bose-Einstein distribution

\BEQ f(\q)=\frac{1}{e^{\beta E(\q)-\alpha}-1},
\EEQ
while the entropy reads

\BEQ \label{s(f)=}
s=k_B\int\frac{\d^3q}{(2\pi\hbar)^3}[-f\log f+(1+f)\log(1+f)].
\EEQ

\section{Decoupled quantum matter in the expanding universe}

In the expanding universe a certain species may decouple from the other matter,
meaning that, given its cross section, the scattering  candidates become so
sparse that practically no scattering will take place anymore. Then in a flat Friedman metric 

\BEQ \label{Friedman}
\d s^2=c^2\d t^2-a^2(t)(\d x^2+\d y^2+\d z^2)
\EEQ
the one-particle
occupation of the mode with momentum $\bf q$ satisfies the free-streaming Boltzmann equation
~\cite{Dodelson,Weinberg,GiuntiKim}

\BEQ \p_tf(q,t)-\frac{\dot a}{a}{\bf q}\cdot \frac{ \p}{\p{\bf q}}f(q,t)=0, \EEQ
with right hand side equal to zero (absence of scattering), and $a$ the scale factor of (\ref{Friedman}).
% in the considered Friedman metric. 
The solution reads
\BEQ \label{fdecp}
f(q,t)=f[a(t)q].\EEQ
we may connect a temperature to this,
%=f\left(\frac{p}{T(t)}\right)=f\left(\beta(t)p\right),\EEQ
\BEQ T(t)=\frac{a_1T_1}{a(t)},\qquad \beta(t)=\frac{1}{k_BT(t)}.
\EEQ
In the approximation of instantaneous decoupling at temperature $T_1$
the distribution function reads %after decoupling

\BEQ\label{fdec}
 f(q,t)=\frac{1}{\exp[\beta_1E(qT_1/T)-\alpha_1]+1},\EEQ
where $\beta_1=1/k_BT_1$ and $\alpha_1=\alpha(T_1)$ are {\it time independent}.
For neutrinos the decoupling took place when they were relativistic, $k_BT_1\gg mc^2$. 
In that case we $\beta_1E(qT_1/T)\approx\beta qc$, so $f$  simplifies to

\BEQ \label{fqr=} f=%\frac{1}{\exp[\beta_1 (T_1pc/T)-\alpha_1]+1}=
\frac{1}{e^{\beta pc-\alpha}+1},
\EEQ
where $\alpha\equiv\alpha_1$ is a constant. Thus even though neutrinos have a mass, their density will be quasi-relativistic, at least 
as long as their distribution is uniform. (It has been argued that neutrinos are presently condensed on matter concentrations such as
 galaxy clusters~\cite{Nnu09}.) 

It has long been supposed that in this non-equilibrium situation thermodynamics would still apply.

With $f$ from (\ref{fqr=}) or, more generally from (\ref{fdecp}), the entropy density (\ref{s(f)=}) is continuous at $T_1$ and scales
as $T^3$ after that, as does the particle density (\ref{n=}), so that the entropy per particle

\BEQ \sigma=\frac{s(t)}{n(t)}=\frac{s(T_1)}{n(T_1)}
\EEQ
is constant, and the entropy in a comoving unit volume $S=N\sigma\sim a^3 s$  conserved.
The same holds for the comoving particle number $N\sim a^3n$.

The energy density is thus simply

\BEA 
%\rho c^2=%\int \frac{\d^3q}{(2\pi\hbar)^3} f(q)
%u=\int\d n(q)E(q)=T^3\int\d n(q)E(qT)
%\sqrt{q^2c^2+m^2 c^4}
u&=& \int \frac{\d^3q}{(2\pi\hbar)^3} \frac{E(q)}{e^{\beta q c-\alpha}+1}\\&=&
\left(\frac{k_B T}{2\pi \hbar c}\right)^3\int\frac{\d ^3k}{e^{k-\alpha}+1}
E\left(\frac{k_BTk}{c}\right). \nn
\EEA
The pressure is not taken from (\ref{pold}) but from (\ref{pnew}), 

%\BEQ p= \int \frac{\d^3q}{(2\pi\hbar)^3} \frac{1}{e^{\beta q c-\alpha}+1}\frac{q^2c^2}{3E(q)}\EEQ
\BEA 
%\rho c^2=%\int \frac{\d^3q}{(2\pi\hbar)^3} f(q)
%u=\int\d n(q)E(q)=T^3\int\d n(q)E(qT)
%\sqrt{q^2c^2+m^2 c^4}
p&=& \int \frac{\d^3q}{(2\pi\hbar)^3} \frac{1}{e^{\beta q c-\alpha}+1}\frac{q^2c^2}{3E(q)}
%\int \frac{\d^3q}{(2\pi\hbar)^3} \frac{E(q)}{e^{\beta q c-\alpha}+1}
\\&=&
\frac{(k_B T)^5}{(2\pi \hbar c)^3}\int\frac{\d ^3k}{e^{k-\alpha}+1}
\frac{k^2}{E({k_BTk}/{c})}. 
\nn
\EEA
This Ansatz has been adopted because it satisfies the energy conservation

\BEQ \dot u+3\frac{\dot a}{a}(u+p)=0.
\EEQ
Now our point is that this is consistent with thermodynamics when taking $s=n\sigma$ from (\ref{s(f)=}) and $\mu$ from 
%the thermodynamic relation
 (\ref{mu=}),

\BEQ \label{mugen}
\mu=\frac{u+p-Ts}{n}=\left\langle E(q)+\frac{q^2c^2}{3E(q)}\right\rangle-T\sigma,
\EEQ
\vspace{0.3cm}

\noindent\noindent
where the averaging is over expression (\ref{fdec}) or (\ref{fqr=}) for $f$.
For $k_BT\ll mc^2$, where $p/u\approx 0$ it yields $\mu\approx mc^2$, which is
consistent with (\ref{dpdT=}) because it cancels the large term $u\approx nmc^2$, 
thus repairing the relation (\ref{thermoprho}).

In conclusion, thermodynamics does explain the non-equilibrium situation of decoupled quantum matter in the expanding Universe, 
but the correspondence $\mu=\alpha k_BT$ between the chemical potential $\mu$  and 
$\alpha$, the time-independent parameter of the Fermi-Dirac distribution (\ref{fqr=}),
is broken and replaced by the general relation (\ref{mugen}).

\vspace{3mm}

\acknowledgements

The first author was inspired by the crystal clear lectures on statistical physics 
by the late B. R. A. Nijboer at the University of Utrecht.

\end{document}